\documentclass[article,pra,twocolumn,nofootinbib]{revtex4-1}

\usepackage{amsmath}
\usepackage{xcolor}
\usepackage{graphicx}
\usepackage{srcltx}
\usepackage{catchfilebetweentags}

\newcommand{\A}{{\bf{A}}}
\newcommand{\B}{{\bf{B}}}

\newcommand{\beq}{\begin{equation}}
\newcommand{\eeq}{\end{equation}}
\newcommand{\ket} [1] {|#1\rangle}
\newcommand{\bra} [1] {\langle#1|}

\newcommand{\ps}{\textbf{ps}}
\newcommand{\CH}{\text{CH}}

\begin{document}

\title{Loss tolerant device-independent quantum key distribution: a proof of principle}


\author{Giuseppe Vallone, Alberto Dall'Arche, Marco Tomasin, Paolo Villoresi}
\address{Department of Information Engineering, University of Padova, I-35131 Padova, Italy}


\begin{abstract}
We here present the rate analysis and a proof of principle realization of a device-independent quantum key distribution (QKD) protocol
requiring the lowest detection efficiency necessary to achieve a secure key compared to device-independent protocols known so far.
The protocol is based on non-maximally entangled state and
its experimental realization has been performed by two-photon bipartite entangled states.
The improvement with respect to protocols involving maximally entangled states has been estimated. 
\end{abstract}

\maketitle

\section{Introduction}
Quantum Key Distribution (QKD) represents an unconditional secure way to share a secret key between two authenticated users,
usually  called Alice and Bob. Photons are the ideal candidates for  QKD implementations due to their low interaction with the environment;
moreover, they can be easily transmitted over long distances with optical fibers \cite{stuc09njp,rose09njp,liu10ope} or free-space links \cite{kurt02nat,hugh02njp,schm07prl,ursi07nap,capr12prl,naue13npho,wang13npho}.
Security of the key is typically proven by using trusted preparation and measurement devices (for a review on QKD security and
experimental implementations, see \cite{scar09rmp}). 
In the last years, great effort have been devoted to the so called Device-Independent QKD (DI-QKD),
aiming at the demonstration of the security when the measuring devices are completely untrusted and their working
mechanism is not known to the users. The key ingredient for DI-QKD is the exploitation of entangled states shared between the
two users: by violating a Bell inequality, it is possible to prove the secrecy of the obtained key.

The DI-QKD offers the advantage that security is
independent of the practical details of the implementation. 
Indeed, the violation of a Bell inequality certifies the
secrecy of the transmission, allowing Alice
and Bob even to use devices directly provided by Eve.
The violation of a Bell inequality without any additional assumption
requires a very high global detection efficiency, from the source to the detectors.
It is well known that non-maximally entangled states offer an advantage, in terms of required detection efficiency,
with respect to maximally entangled state to violate the Clauser-Horne-Shimony-Holt (CHSH) inequality ~\cite{clau74prd,eber93pra}.
Recently, detection loophole-free violations of the CHSH inequality by non-maximally entangled photons
were indeed reported~\cite{gius13nat,chri13prl}.
Non-maximally entangled states were proven to be also useful for several bipartite Bell inequalities  \cite{vall14pra} and 
for quantum steering \cite{vall13pra}.

The ent-B92 protocol, a version of the B92 protocol~\cite{benn92prl} realized
with non-maximally entangled states, was proposed in~\cite{luca12pra}.
We here propose its generalization and analyze its secret key rate
when detection inefficiencies are taken into account.
We moreover present a proof-of-principle realization of the protocol by exploiting non-maximally entangled states produced by 
spontaneous parametric down conversion.
With {\it proof-of-principle} we mean that we demonstrated that it is in principle possible to realize the DI protocol 
with our experimental generated state if higher detection efficiencies were used. 
Due to the efficiency of our setup (about 10\%) it was not possible to achieve a complete DI-QKD demonstration. However, by using eq. (13) 
(see below), we will predict the experimental value of the Bell parameter achievable with our experimentally generated state in case 
of a given efficiency $\eta$. By this prediction we will estimate the achievable rate of our DI-QKD protocols. 

We will show that the protocols here presented allow a DI-QKD
security with the lowest required threshold detection efficiency to date.

\section{Generalized \lowercase{ent}-B92 protocol}

Let's consider Alice and Bob sharing the following non-maximally entangled state:
\begin{align}
\label{eq:entangled} |\Phi(\theta)\rangle_{AB} =
\cos\frac\theta2{\ket{H}}_{A}{\ket{H}}_{B}+\sin\frac\theta2{\ket{V}}_{A}{\ket{V}}_{B}\,,
\end{align}
where $\ket{H}$ and $\ket{V}$ are the horizontal and vertical polarization states and $0<\theta \leq\pi /2$. 
The parameter $\theta$ is monotonically related to the amount of entanglement, since 
the concurrence \cite{woot98prl} is given by $C = \sin\theta$.

The protocol works as follow: 
Alice measures with low probability $p\ll1$ 
its photon along the $\mathcal A_1\equiv\{\ket{a_1},\ket{\bar a_1}\}$ basis, with
$\ket{a_1}=\ket{V}$ and $\ket{\bar a_1}=\ket{H}$.
With high probability $1-p$ she measures along the 
 $\mathcal A_0\equiv\{\ket{a_0},\ket{\bar a_0}\}$ basis, 
where $\ket{a_0}=\frac{1}{\sqrt2}(\ket H+\ket V)$ and $\ket{\bar a_0}$ is its orthogonal state;
Bob randomly and with probability $1/2$ measures the incoming states in the
$\mathcal B_0$ or $\mathcal B_1$ basis where 
$\mathcal B_k = \{ |b_k\rangle, |\overline{b}_k\rangle \}$ 
and
\beq\label{measure}
\begin{aligned}
\ket{b_k}
&=\sin\frac{\varphi}2\ket{H}-(-1)^{k}\cos\frac{\varphi}2\ket{V}\,,
\\
\ket{\bar b_k}
& = \cos\frac{\varphi}2 \ket{H} +( -1)^{k}\sin\frac{\varphi}2\ket{V}\,.
 \end{aligned}
\eeq
The results from Alice's $\mathcal{A}_0$ basis
measurements are used as bits of the raw key together with Bob's results, while those
from the $\mathcal{A}_1$ basis will be used to perform a
test against the eavesdropper attack, as in the uninformative states B92 QKD protocol (us-B92) introduced in ~\cite{luca09pra}.
On Alice side, the states $\ket{a_0}$ and $\ket{\bar a_0}$ correspond to bits 0 and 1 respectively.
Upon obtaining the state $\ket{b_k}$ 
Bob decodes Alice's bit as $j=k\oplus1$ (the symbol $\oplus$ means
``addition modulo 2'') and labels the result as
\textit{conclusive}; on the contrary, upon obtaining the state
$|\bar b_k\rangle$, Bob labels the result as
\textit{inconclusive}. The probability of a conclusive event is given by $P_c=\frac12(1-\cos\theta\cos\varphi)$,
independent on Alice measurements.
The {\it sifted key} is obtained by selecting the conclusive results
corresponding to Alice's $\mathcal{A}_0$ measurements.
The ent-B92 protocol of~\cite{luca12pra} corresponds to the choice $\varphi=\theta$.

The main problem of a fully DI-QKD protocol is related to the so called {\it detection loophole}, namely the 
fact that the photon detection is inefficient and, if the detectors are not trusted, 
an eavesdropper can exploit this inefficiency to gain information on the key.
In the next section we will show how to extract a secure key in the device-independent scenario
in presence of detection (and transmission) inefficiencies.

\section{Key Rate Analysis} \label{sec:rate_analysis}
In this section we will derive the secret key rate (see equation \eqref{final_rate}) of the generalized ent-B92 protocol
in case of detection inefficiencies, improving the results obtained in \cite{luca12pra}.

Let's consider a transmission of $\mathcal N$ pairs in which Alice chooses the $\mathcal A_0$ basis and Bob chooses
with probability $\frac12$ the basis $\mathcal B_0$ or $\mathcal B_1$.
The $\pm 1$ outputs of Alice's measurements correspond to bits 1 and 0 of the sifted key.
The overall efficiencies (including transmission and detection efficiencies)
are given by $\eta_A$ and $\eta_B$ and Alice and Bob must decide a strategy to deal with non-detection events.
On Bob's side, only $+1$ outputs, the so called conclusive outcomes, are taken into account to build the key: thus, on Bob side,
non-detection event will be associated to $-1$ output (corresponding to non-conclusive outcomes).
Then, all the Alice's bits corresponding to non-conclusive Bob outcomes, can be simply discarded as it is usually
done in the sifting phase of the BB84 protocol.
On the other side, when Alice measures in the $\mathcal A_0$ basis and does not obtain physical detection, she randomly
chooses an outcome: whatever value
she decides to output, the bit will enter into the key. Alice thus assigns to non-detection events the $\ket{a_0}$ or $\ket{\bar a_0}$
outcome with $1/2$ probability in this case.

Due to inefficiencies, Bob receives $\mathcal NP_c\eta_B$ conclusive counts. 
After the sifting phase, Alice string $\A^c$ and Bob's string $\B^c$ consists of $\mathcal NP_c\eta_B$ bits.
In Ref \cite{luca12pra} (see in particular eq. (11) and (12)), the secure key rate -- 
the ratio between secure bits and the overall sent pairs 
-- is given 
 in term of the quantum bit error rate $Q^c$ and
the Clauser-Horne (CH) parameter $S_\CH = P(a_1,b_1)+P(a_0,b_1)+P(a_1,b_0)- P(a_0,b_0)-P(a_1)-P(b_1)$~\cite{clau74prd} as
\beq\label{old_rate}
\tilde r=\eta_A\eta_BP_c\left[1-h_2(Q^c)-\log_2f(S_\CH)\right]\,,
\eeq
where
\beq
 f(S_\CH) =1+\sqrt{1-4{ S_\CH}-4{S^2_\CH}}\,.
\eeq
and $h_2(x)$ is the binary entropy given by $h_2(x)=-x\log_2x-(1-x)\log_2(1-x)$.
The rate is derived under the  assumption that the measurement devices are memoryless.
In the previous expression $P(a_i,b_j)$ is the joint probability that Alice measures the
state $\vert a_i\rangle$ and Bob detects the state $\vert
b_i\rangle$, while $P(a_1)$ and $P(b_1)$ are the probabilities
that Alice and Bob respectively measure $\ket{a_1}$ and
$\ket{b_1}$, regardless of what is measured by the other user. The quantum bit error rate (QBER) $Q^c$,
defined as the ratio of the number of errors over the number of conclusive outcomes, must be evaluated over the sifted strings
$\A^c$ and $\B^c$.

By using a technique introduced in \cite{bran12pra},
 Alice and Bob can improve the secure key rate: they
 will perform a post-selection on $\A^c$ and $\B^c$,
 by selecting only the bits in which also Alice obtained a physical detection.
 We called $\A^\ps$ and $\B^\ps$ the post-selected Alice's and Bob's strings, 
with length  $\mathcal NP_c\eta_A\eta_B$.

The length of the secure key can be bounded by \cite{ren12ieee}
\beq
\ell\geq H_{min}(\B^\ps|E)-H(\B^\ps|\A^\ps)\,,
\eeq
where $H_{min}(\B^\ps|E)$ is the min-entropy of $\B^\ps$ conditioned on Eve's information.
As usual $H(\B^\ps|\A^\ps)=\mathcal NP_c\eta_B\eta_Ah_2(Q^\ps)$
is related to the classical error correction protocol between the $\A^\ps$ and $\B^\ps$ strings.
In the previous expression $Q^\ps$ is the QBER on the post-selected data, and, its theoretical value $Q^\ps_{th}$
in the case of no channel or measurement errors, is given by :

\beq
Q^\ps_{th}=\frac{1-\cos(\theta -\varphi)}{2-2 \cos\theta \cos\varphi}\,.
\eeq
The choice $\varphi=\theta$, used in the ent-B92 protocol, gives null QBER.
We will see that this choice doesn't always represent the optimal choice for the secure key rate.

As demonstrated in \cite{bran12pra}, by using the chain rule and the data-processing inequality for smooth min-entropy
\cite{toma10ieee,ren12ieee}, it is possible to bound  Eve's information on the sifted bits by using her information on $\B^c$:
\beq
H_{min}(\B^\ps|E)\geq H_{min}(\B^c|E)-\mathcal NP_c\eta_B(1-\eta_A)\,,
\eeq
where $\mathcal NP_c\eta_B(1-\eta_A)$ is the difference between the $\B^c$ and the $\B^\ps$ string length.

As shown in \cite{koni09ieee}, the min-entropy can be related to the maximal probability of guessing the key bits, namely
\beq
H_{min}(\B^c|E)=-\mathcal NP_c\eta_B\log_2P_{\text{guess}}({\bf b}|E)\,.
\eeq
By using the results of \cite{masa11nco}, the probability of guessing the bits can be related to the Bell inequality by
$P_{\text{guess}}({\bf b}|E)\leq
\frac{ f(S_\CH)}{2}$.
The final secure key length can be thus written as
$\ell\geq \mathcal NP_c\eta_B\left[ \eta_A(1-h(Q^{\textbf{ps}}))-\log_2f(S_\CH)\right]$
and the final rate $r=\ell/\mathcal N$ becomes
\beq\label{final_rate}
r=\eta_BP_c\left[\eta_A(1-h_2(Q^{\textbf{ps}}))-\log_2f(S_\CH)\right]\,.
\eeq
As usual, in the secure rate formula, the $h_2(Q^\ps)$ 
term corresponds to the bits used for error correction, while the
$\log_2$ contribution is related to Eve's knowledge on the key and the required compression in the privacy amplification
stage. 
The violation of the CH inequality $S_\CH\leq0$~\cite{clau74prd} is a test against the local-realism of quantum physics:
it can be trivially checked that, when the inequality is not violated, the secure rate \eqref{final_rate} is zero\footnote{If the
key rate $r$ obtained in \eqref{final_rate} is negative, no secure key can be distilled.}.

In case of perfect efficiencies, the theoretical value of $S_\CH$ for the non-maximally entangled state \eqref{eq:entangled} and the measurement defined in \eqref{measure}
is given by $S_{\text{CH}}(\theta,\varphi)=\frac{1}{2} (\cos\varphi+\sin\theta \sin\varphi-1)$.
The choice $\varphi=\arctan(\sin\theta)$
leads to the maximum achievable violation  with
the state \eqref{eq:entangled}, namely $S^{max}_\CH(\theta) = \frac12(\sqrt{\sin^2\theta + 1} -1)$.
\begin{figure}[t]
\centering\includegraphics[width=8.8cm]{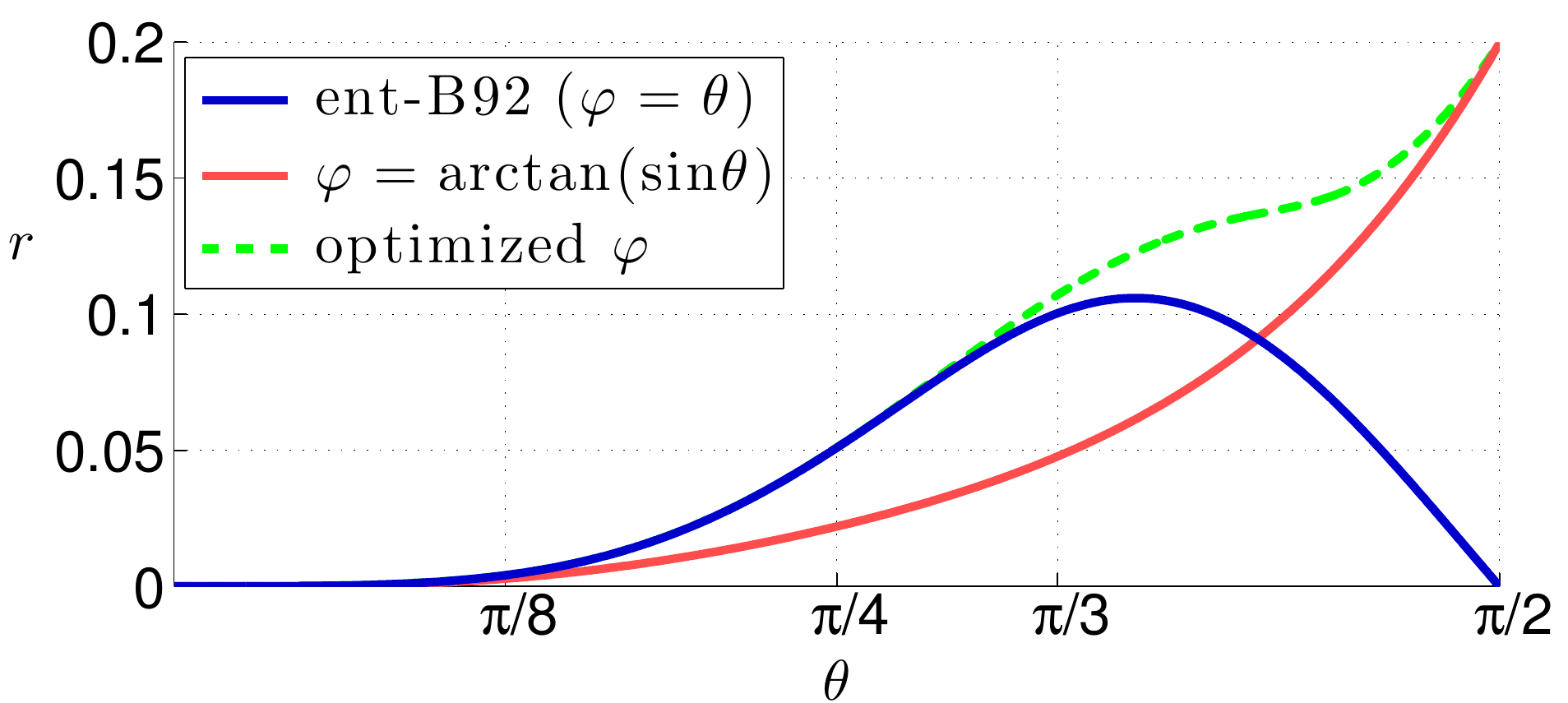}
\caption{Theoretical secure key rate $r$ for the generalized ent-B92 protocol, in case of perfect detection efficiencies.}
\label{fig:th-rates}
\end{figure}
It is worth noting that considering a trusted measurement device is equivalent to taking perfect efficiencies,
namely $\eta_{A}=1$ and/or  $\eta_{B}=1$. 
In fact, if the device is trusted, we can safely consider only the detected events.
In this way, we can have three possible scenarios: 
 full DI-QKD when the actual efficiencies are considered, one-side device independent-QKD (1SDI-QKD)\cite{bran12pra}
in which only one of the two devices
(Alice or Bob) is trusted, and standard QKD with both trusted devices. 
In case of standard QKD (corresponding to $\eta_A=\eta_B=1$), the achievable  rate with the ent-B92 protocol (corresponding to $\varphi=\theta$),
is shown in Fig.~\ref{fig:th-rates} with the maximum rate obtained for $\theta\simeq65.28^\circ$.
By using the angle that maximizes the violation of the Bell inequality ($\varphi = \arctan(\sin\theta)$) 
it is possible to improve the rate when $\theta\gtrsim71.62^\circ$  (see Fig.~\ref{fig:th-rates}).
More generally, it is possible to numerically optimize the value of the parameter $\varphi=\varphi^\star(\theta)$
in function of $\theta$ to maximize the achievable rate, as shown with dashed line in Fig.~\ref{fig:th-rates}. 
Note that, whenever $\varphi\neq \theta$, the theoretical QBER is not vanishing:
however, the non vanishing QBER can be compensated by a larger violation of the CH inequality, allowing more
secrecy in the privacy amplification stage. 
It is clear that, when Alice and Bob have trusted devices
(corresponding to the fair sampling assumption of non-locality tests),
the ent-B92 protocol cannot offer advantages with respect to the entangled version of the BB84 protocol 
\cite{eker91prl, masa11nco}. In fact, in this case, the secure key rate of ent-B92 is always lower than the BB84,
given by $r_{BB84}=1-2h_2(Q)$.
As we will see, the advantages come when one (or both) device is not trusted: in this case,
the lower threshold detection required by the (generalized) ent-B92 protocol to violate the CH inequality, 
gives considerable improvement of the secure key rate with respect to protocols based on maximally entangled states.

The rate achieved in \eqref{final_rate} on the post-selected data can be compared to the one obtained without post-selection \eqref{old_rate},
with QBER $Q^c$ evaluated over all conclusive events. Since Alices  assigns to non-detection events the $\ket{a_1}$ or $\ket{\bar a_1}$
outcomes with $1/2$ probability, in this case the QBER of the sifted key will be
\beq
Q^c=\eta_AQ^{\textbf{ps}}+\frac{1-\eta_A}{2}\,.
\eeq
It is easy to show that the rate $\tilde r$ \eqref{old_rate} is lower than the rate $r$ \eqref{final_rate} achievable with the post-selection technique
for any $\eta_A<1$ while $\tilde r=r$ when $\eta_A=1$.

It is also useful to compare the result obtained in \eqref{final_rate} 
to the one obtained with the post-selection technique in \cite{bran12pra} using the usual DI-QKD protocol of \cite{masa11nco} implemented with maximally entangled states (the entangled version of the BB84 protocol):
\beq
r'=\eta_A\eta_B(1-h_2(Q^{\textbf{ps}}))-\log_2f(S_\CH)\,.
\eeq
The difference from $r$ and $r'$ arises from the fact that in the BB84 protocol
the key is obtained by using the results of Bob in a single basis,
while in the generalized ent-B92 protocol the key is obtained by keeping the 
Bob's conclusive results in the basis $\mathcal B_0$ and $\mathcal B_1$.

\begin{figure}[t]
\centering\includegraphics[width=8.8cm]{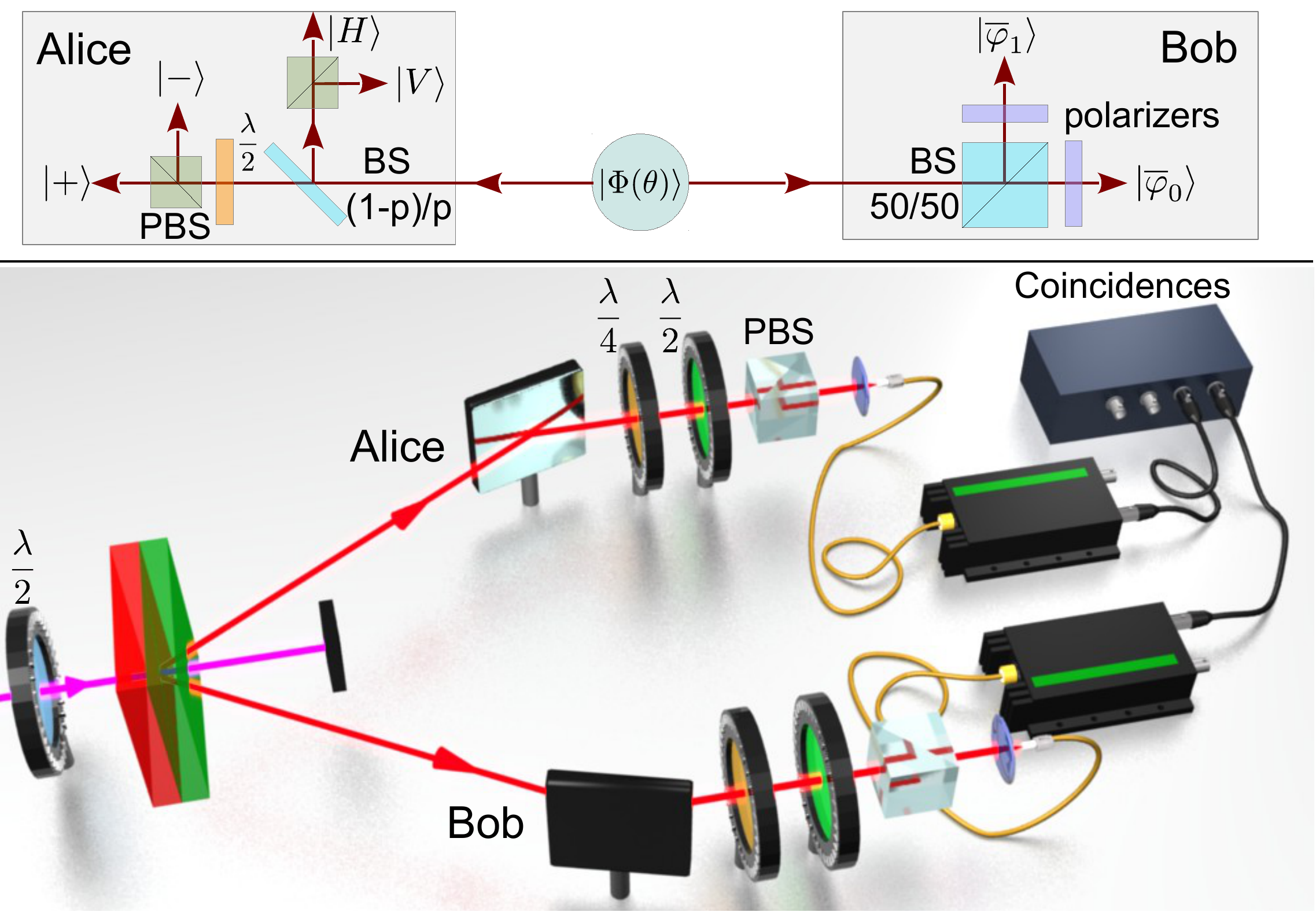}
\caption{(top) Scheme of the generalized ent-B92 scheme. (bottom) 
Experimental setup used for the generation of the non-maximally entangled state and the realization of the protocol.}
\label{setup}
\end{figure}

\section{Proof of principle realization}

In this section we present a proof of principle realization of the protocols above described.
As anticipated in the introduction, we will not present a complete DI-QKD demonstration
due to the low detection efficiencies measured in our setup.
By generating a two-photon non-maximally entangled state by 
spontaneous parametric down conversion, we
will demonstrate that it is in principle possible to realize the DI protocol 
with our experimental generated state if higher detection efficiencies were used.
We first describe the results  obtained in a trusted scenario (standard QKD) in order to test our entanglement source
and then analyze what can be achieved in a DI framework.

Our source of non-maximally entangled states, shown in Fig. \ref{setup},
 is given by two overlapped Type-I non-linear BBO crystals shined by a pulsed UV laser
at $405nm$. The two spontaneous parametric down converted photons are emitted at $810nm$.
The two crystals have the optical axis rotated by $90^\circ$: the first crystal generates the $\ket{HH}$ pairs, while the second crystal
generates the $\ket{VV}$ pairs. By using a pump laser with polarization $\cos\frac\theta2\ket V_{p}+\sin\frac\theta2\ket H_{p}$
the non-maximally entangled state \eqref{eq:entangled} can be generated. By varying the linear polarization on the pump laser
it is possible to change the relative contribution of the $\ket{HH}$ and $\ket{VV}$  terms in the generated state.
The UV laser has pulse duration of about $10ps$ and 76MHz repetition rate. Due to the long coherence time of the pump laser,
it is not necessary to compensate the temporal walk-off in the BBO crystals.
We note that, by this protocols, Bob does not need to project into the $\ket{\overline b_k}$
states, since they don't appear in the CH inequality neither are  used in the sifted key generation. Bob measurements can be thus
restricted to the positive-operator valued measure (POVM) $\Pi_0=\frac12\ket{b_0}\bra{b_0}$, 
$\Pi_1=\frac12\ket{b_1}\bra{b_1}$ and $\Pi_{inconc}=\openone- \Pi_0-\Pi_1$.
From the experimental point of view the Bob measurement can be simply represented by a beam splitter and two polarizers.

\begin{figure}[t]
\centering\includegraphics[height=4.3cm]{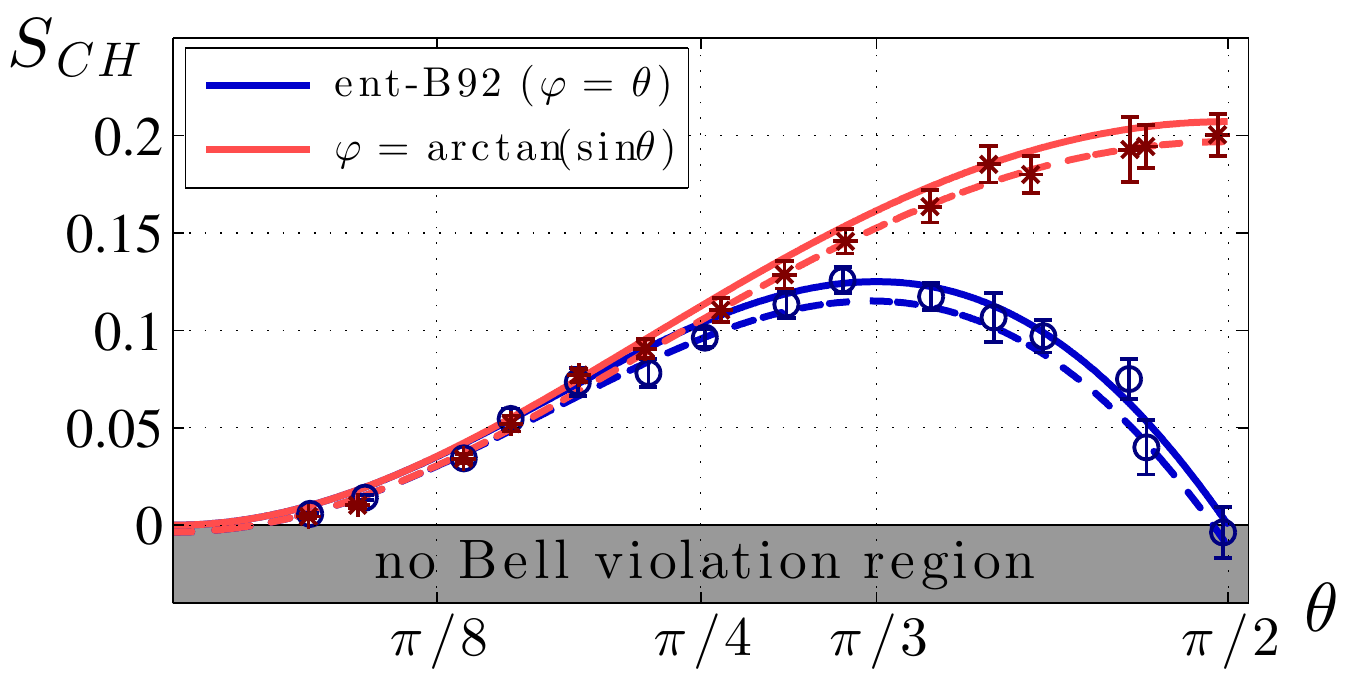}
\includegraphics[height=4.3cm]{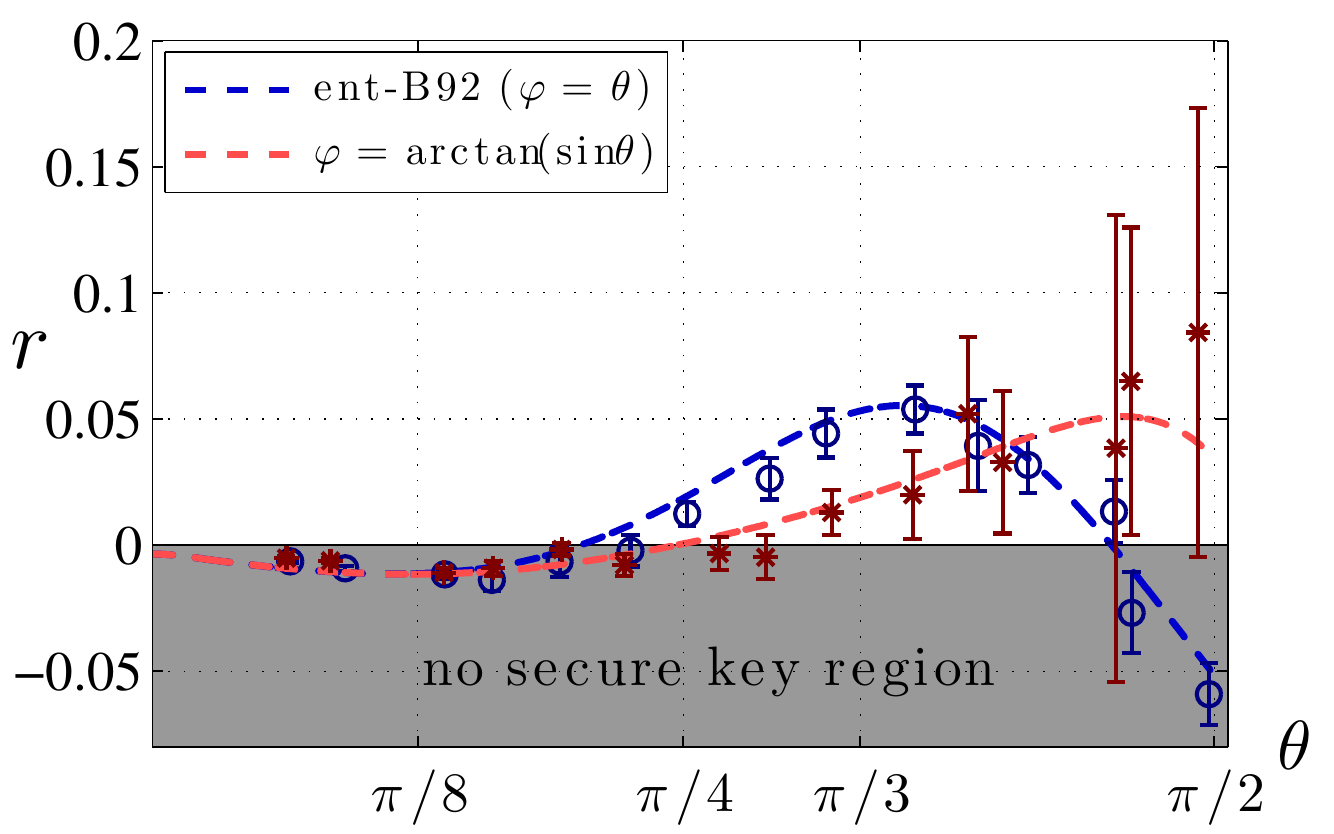}
\caption{(left) Experimental value of the parameter $S_\CH$ and corresponding errors for the ent-B92 (blu circles) and the
$\varphi=\arctan(\sin\theta)$ protocol (red stars). 
(right) Experimental key rates for the two protocols with trusted measurement devices. 
Continuous and dashed lines refer to theoretical predictions
 corresponding to perfect state generation and noise model \eqref{model} respectively. }
\label{fig:exp-rates}
\end{figure}

As said, we tested our source by measuring the Bell parameter $S_\text{CH}$ and the achievable secure key rate  
in case of trusted measurement devices.
In figure \ref{fig:exp-rates} we show the experimental value of $S_\text{CH}$ and the secure key rate $r$
 in function of the entanglement parameter $\theta$.
If the obtained rate $r$ is below zero, no secure key can be extracted. In order to take into account
imperfections in the setup, the experimental
generated state can be expressed by the following noise model:
\beq\label{model}
\rho_{exp}=(1-p_c-p_w)\ket{\Phi(\theta)}\bra{\Phi(\theta)}+p_c\rho_c+p_w\frac{\openone}{4}	\,.
\eeq
In the previous equation the state $\rho_c$ is given by
$\rho_c=\cos^2\frac\theta2\ket{HH}\bra{HH}+\sin^2\frac\theta2\ket{VV}\bra{VV}$
and $p_c$ ($p_w$) represent the amount of colored (white) noise.
Dashed line in Fig. \ref{fig:exp-rates} represent the theoretical value of the $S_\text{CH}$ parameter and the secure rate $r$
obtained by the state \eqref{model} with $p_w=0.007$ and $p_c=0.015$, with good agreement between the model
and the obtained results. To further validate our noise model we present in  \ref{error-model} its prediction 
 for the threshold detection efficiencies required for the violation of the Bell inequality.

 Let us now analyze what can be achieved in a DI framework. We calculated the overall detection efficiency of our system
 by evaluating the ratio between the measured coincidences and the single counts, thus taking into account transmission, 
  coupling into fibers and detection losses. 
  Since the measured detection efficiency is about $10\%$ for both Alice and Bob we cannot achieve DI secure key rate.
Nevertheless, we can estimate the key rate achievable with given detection efficiencies $\eta_A$ and $\eta_B$.
Indeed, by assuming that probabilities $P(a_i,b_j)$ are related to the probabilities $p(a_ib_j)$ normalized on the 
post-selected events in which Alice and Bob have a coincidence,
we can predict the CH parameter as (see \ref{predicted_SCH} for its derivation):
\begin{widetext}
\begin{align}
\label{SCHexp}
\notag
S_\CH=&\eta_A\eta_B[p(a_1b_1)+\frac12p(a_0b_1)+p(a_1b_0)-\frac12p(a_0b_0)+\frac12p(\bar a_0b_0)-\frac12p(\bar a_0b_1)]
\\
&-\eta_A[p(a_1b_0)+p(a_1\bar b_0)]-\frac{\eta_B}2[p(a_0b_1)+p(\bar a_0 b_1)+p(a_0b_0)+p(\bar a_0 b_0)]
\end{align}
\end{widetext}

The previous expression can be derived under the condition that
on the $\mathcal A_1$, $\mathcal B_0$ and $\mathcal B_1$ basis, non-detection events are associated to $-1$ outputs,
namely to the states $\ket{\bar a_1}$, $\ket{\bar b_0}$ and $\ket{\bar b_1}$
while the states  $\ket{a_0}$ and $\ket{\bar a_0}$ are randomly chosen in case of non-detection in the $\mathcal A_0$ basis.

\begin{figure}[t]
\centering\includegraphics[height=3.5cm]{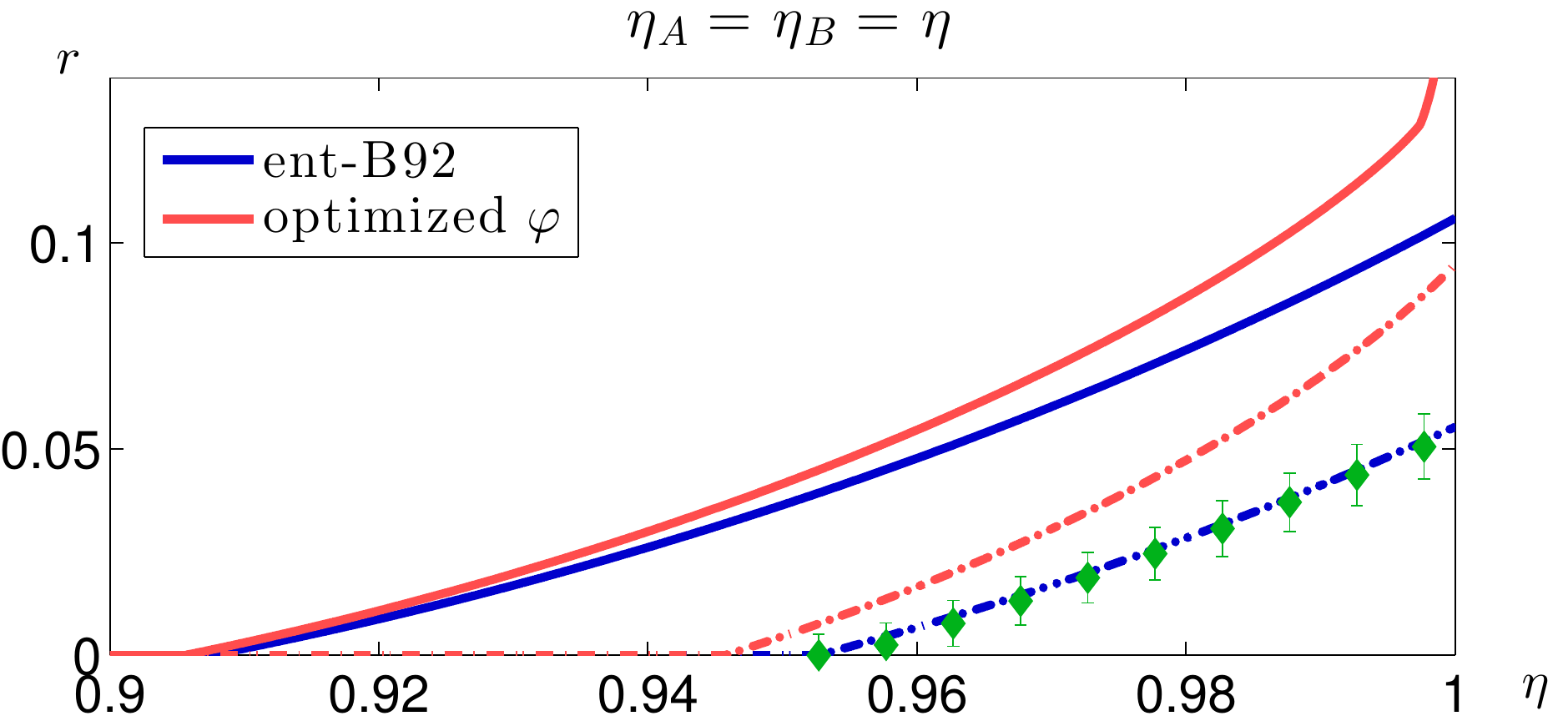}
\includegraphics[height=3.5cm]{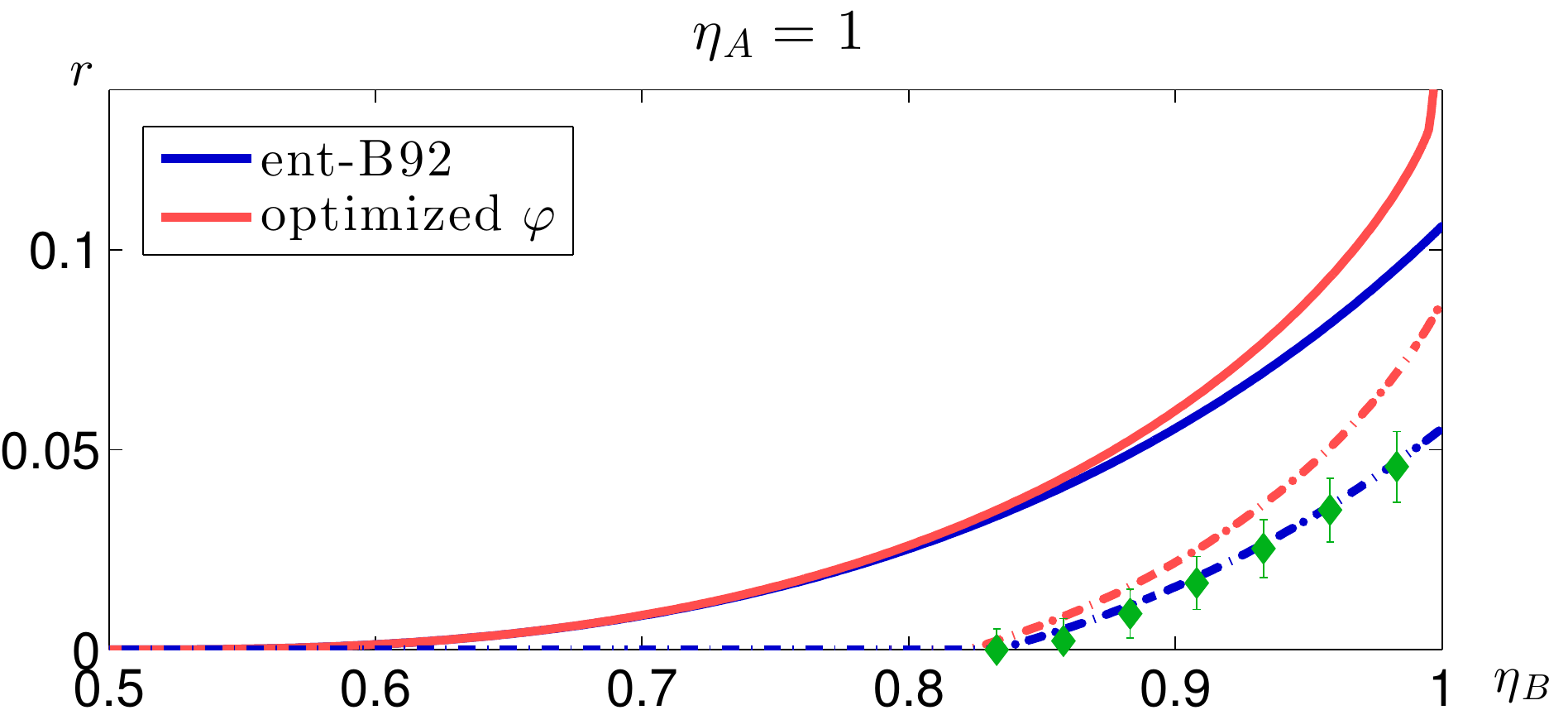}
\caption{Key rates achievable for (left) the full DI-QKD case ($\eta_A=\eta_B=\eta$) and for (right)
the 1SDI-QKD ($\eta_A=1$) in function of the detection efficiency. 
For 1SDI-QKD the rate is the the amount of secure bits over the detected Alice bits.
With continuous lines we indicate the theoretical rates achievable with the non-maximally entangled state
$\ket{\Phi(\theta)}$. With dashed lines we indicate the predicted rates obtained by using the noise model of eq. \eqref{model}, while
with green squares we indicate the rates achievable when the experimentally measured probabilities 
are used to evaluate eq. \eqref{SCHexp}.
}
\label{fig:reta}
\end{figure}

Then, by measuring the probabilities appearing in equation \eqref{SCHexp}, 
we can estimate the value of the Bell parameter in case of arbitrary efficiencies $\eta_A$ and $\eta_B$ and
thus predict the secure key rate achievable with our experiment states when more efficient detectors are used.  
It is worth noticing that overall efficiencies of the order of 75\% were already demonstrated in the lab 
by using superconducting TES detectors in \cite{gius13nat, chri13prl}. 

Let's first consider full DI-QKD with Alice and Bob having the same efficiency $\eta_A=\eta_B=\eta$.
For each value of $\eta$ is possible to optimize the value of $\theta$ (or $\theta$ and $\varphi$) that maximizes the key rate for the ent-B92 (or generalized ent-B92) protocol: in figure \ref{fig:reta}(left) we illustrate the achievable key rate in function of the detection efficiency for the ent-B92 and the generalized
protocol.
We note that positive secure key rate can be obtained up to an efficiency of 90.57\%, improving the results of $90.9\%$ and $91.1\%$ obtained respectively in \cite{ma12qic} and \cite{bran12pra}. We also show the rates that can be achieved with our experimental generated state. 
In particular,  we indicate with dashed lines the predicted rates achievable by using the theoretical noisy state of eq. \eqref{model}
to calculate the probabilities appearing in eq. \eqref{SCHexp}. With green squares we report the rates achievable when the probabilities 
that we have experimentally measured are used to evaluate eq. \eqref{SCHexp}.
Such rates are below the rates achievable with a perfect non-maximally entangled state $|\Phi(\theta)\rangle$
due to the presence of decoherence and white noise causing a decrease of the state purity. 
Our measurements indicate that it is necessary to generate states with high purity in order to reach low detection efficiencies.

Great improvement with respect to state-of-the-art results are obtained by considering one-side DI-QKD in which Alice device is trusted, 
corresponding to $\eta_A=1$ in the secure key rate \eqref{final_rate} and in the predicted Bell parameter \eqref{SCHexp}.
In this case the rate $r$ correspond to the fraction of secure bits over the Alice detected bits.  
In figure \ref{fig:reta}(right), we show the achievable key rate in function of the detection efficiency in the one-side DI-QKD case. 
For the ent-B92 protocol, the secure key rate (without experimental imperfection)
becomes $r=\eta_BP_c\left[1-\log_2f(S_\CH)\right]$, which is positive whenever the Bell inequality $S_{CH}\leq0$ is violated.
Note that the same rate can be obtained by the original analysis of the ent-B92 protocol \cite{luca12pra} applied for the one-side case.
We remark that it is possible to obtain positive secure key rate  up to a detection efficiency of $50\%$, 
improving the result obtained in \cite{bran12pra}
in which an efficiency greater than $65.9\%$ is required for key generation. 
Also in this case we show the rate predicted by our experimental data: these data shows again that,
in order to fully exploit the properties of low entangled states, it is necessary to generate quantum state with low amount of noise.

Our result closes the gap, from the theoretical point of view, 
between one-side Bell inequality (also known as steering inequality \cite{jone07pra,wise07prl,
bran12pra}) and key generation in 1SDI-QKD, 
since in our protocol the violation of the Bell inequality corresponds
to a positive secure key rate. 
For fully DI-QKD still remains a gap, which is due to difference between the threshold of $\eta > 82.8\%$ needed
for a violation of the CHSH inequality \cite{eber93pra} and the efficiency required for the security of DI-QKD, namely $\eta > 90.57\%$.

\section{Conclusions} 
We have derived an efficient key rate for the (generalized) ent-B92 protocol in case of detection
inefficiencies. We experimentally tested our result with two-photon non-maximally entangled states with good agreement
between theory and experiment. The protocol 
is able to achieve secure key rate with the lowest  detection efficiency to date. 
While the improvement for the full DI-QKD case is small and has mainly theoretical relevance (we lowered the threshold efficiency from $90.9\%$ to $90.57\%$),
great improvement is obtained in the one-side DI-QKD: it is possible to achieve positive secure key rate up to $50\%$ efficiency,
to be compared with the state of the art result of $65.9\%$ \cite{bran12pra}.

\section*{Acknowledgments}
Authors would like to thank Prof. P. Mataloni from the University Sapienza of Roma (IT), 
Dr. G. Di Giuseppe from University of Camerino (IT) and Dr. M. Lucamarini from Toshiba Research Europe Ltd (UK)
for useful and stimulating discussions.
Our work was supported by the Strategic-Research-Project QUANTUMFUTURE (STPD08ZXSJ) of the University of
Padova and Strategic-Research-Project QUINTET of the Department of
Information Engineering, University of Padova.


\clearpage

\onecolumngrid
\appendix

\section{Predicted Bell parameter in case of detection inefficiencies}
\label{predicted_SCH}
Let's consider the Clauser-Horne (CH) parameter 
\beq\label{original_SCH}
S_\CH = P(a_1,b_1)+P(a_0,b_1)+P(a_1,b_0)- P(a_0,b_0)-P(a_1)-P(b_1)\,,
\eeq
with the probabilities $P$'s  normalized over all generated pairs. In case of detection inefficiencies $\eta_A$ and $\eta_B$ we can predict
the values of each probability. 
In the generalized ent-B92 protocol, non-detection events are associated to the states $\ket{\bar a_1}$, $\ket{\bar b_0}$ and $\ket{\bar b_1}$
when the observables  $\mathcal A_1$, $\mathcal B_0$ and $\mathcal B_1$ are respectively measured. 
Then, the probabilities $P$'s can be predicted as 
\beq
\label{Pa1}
\begin{aligned}
P(a_1,b_1)&=\eta_A\eta_Bp(a_1b_1)\,,\quad &P(a_1)&=\eta_Ap(a_1)=\eta_A[p(a_1b_0)+p(a_1\bar b_0)]\,,
\\
P(a_1,b_0)&=\eta_A\eta_Bp(a_1b_0)\,, &P(b_1)&=\eta_Bp(b_1)=\eta_B[p(a_0b_1)+p(\bar a_0b_1)]\,,
\end{aligned}
\eeq
where $p(a_ib_j)$ are the probabilities normalized on the post-selected events in which Alice and Bob have a coincidence.

On the other side, when measuring the $\mathcal A_0$ observable, it is necessary to remember that,
 in the case of non-detection, the state  $\ket{a_0}$ is chosen with probability $\frac12$.
 Then 
 \beq
 \label{Pa0}
\begin{aligned}
 P(a_0,b_0)&=\eta_A\eta_Bp(a_0b_0)+(1-\eta_A)\eta_B\frac12p(b_0)
 \\
 &= \eta_A\eta_Bp(a_0b_0)+(1-\eta_A)\eta_B\frac12[p(a_0b_0)+p(\bar a_0b_0)]\,,
\\
 P(a_0,b_1)&=\eta_A\eta_Bp(a_0b_1)+(1-\eta_A)\eta_B\frac12p(b_1)
 \\
 &= \eta_A\eta_Bp(a_0b_1)+(1-\eta_A)\eta_B\frac12 [p(a_0b_1)+p(\bar a_0b_1)]\,.
 \end{aligned}
\eeq
By inserting equation \eqref{Pa1} and \eqref{Pa0} into \eqref{original_SCH} we obtained the predicted Bell parameter as
\begin{align}
\notag
S_\CH=&\eta_A\eta_B[p(a_1b_1)+\frac12p(a_0b_1)+p(a_1b_0)-\frac12p(a_0b_0)+\frac12p(\bar a_0b_0)-\frac12p(\bar a_0b_1)]
\\
&-\eta_A[p(a_1b_0)+p(a_1\bar b_0)]-\frac{\eta_B}2[p(a_0b_1)+p(\bar a_0 b_1)+p(a_0b_0)+p(\bar a_0 b_0)]\,.
\end{align}

\section{Error model verification}
\label{error-model}
As written in the main text, we used a noise model in order to take into account the possible noise sources in our setup. 
The noise model takes into account two possible sources: the first is due to the effect of background photons, 
modeled by the addition of a white noise $\frac{\openone}4$ 
and corresponding to a depolarizing channel. The second is due to a partial distinguishability of the $\ket{HH}$
and $\ket{VV}$ events generated by the two SPDC crystals causing partial decoherence: this effect is modeled by a colored noise $\rho_c$,
corresponding to a phase damping channel \cite{Nielsen&Chuang}.
Therefore, the generated state can be expressed by the following model:
\beq\label{model_appendix}
\rho_{exp}=(1-p_c-p_w)\ket{\Phi(\theta)}\bra{\Phi(\theta)}+p_c\rho_c+p_w\frac{\openone}{4}\,.
\eeq
In the previous equation 
$\rho_c=\cos^2\frac\theta2\ket{HH}\bra{HH}+\sin^2\frac\theta2\ket{VV}\bra{VV}$
and $p_c$ ($p_w$) represents the amount of colored (white) noise.
To further validate our noise model we checked the predictions of our model with the experimental data for the threshold detection efficiencies.
\begin{figure}[t]
\includegraphics[width=8cm]{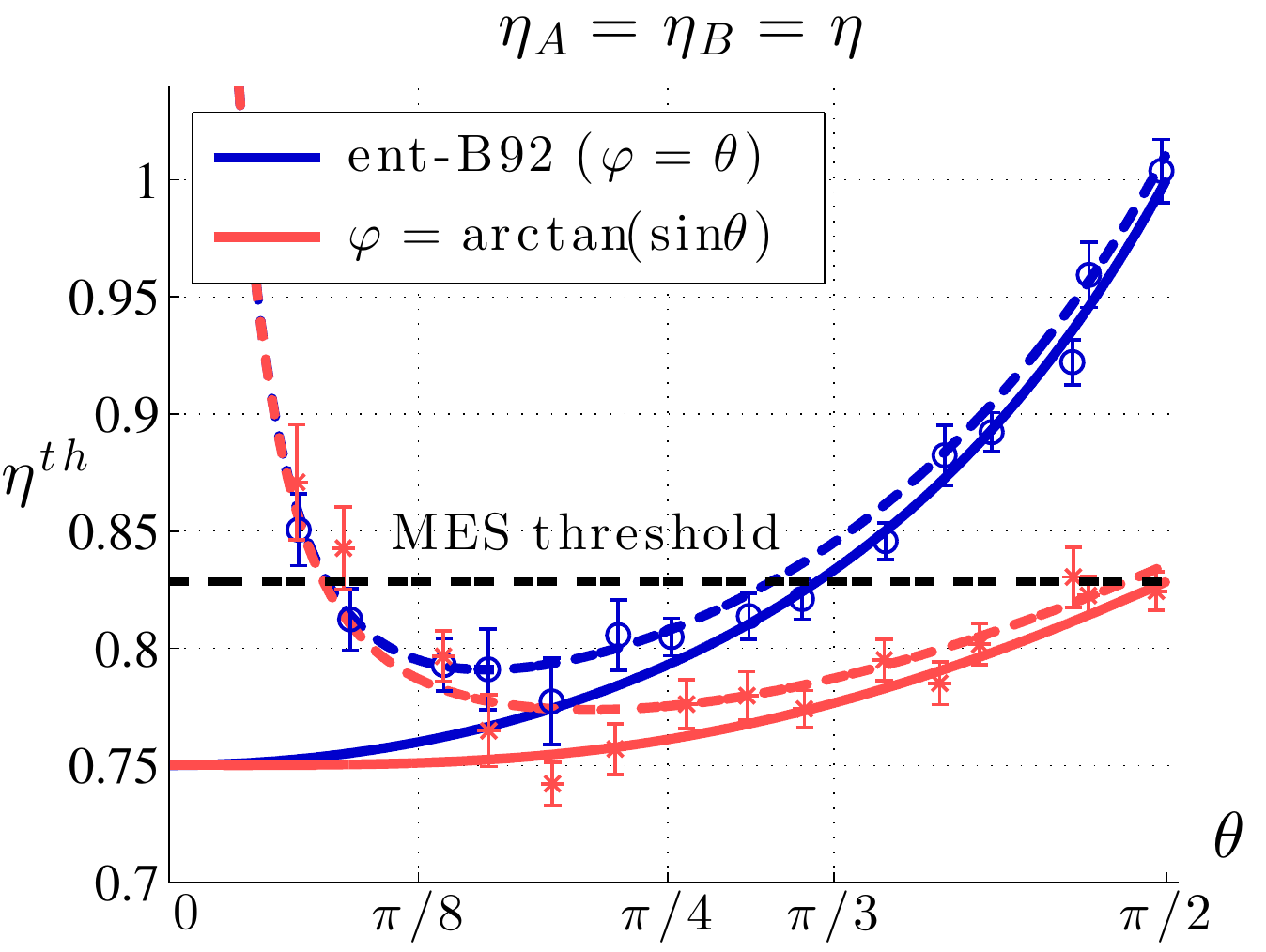}
\includegraphics[width=8cm]{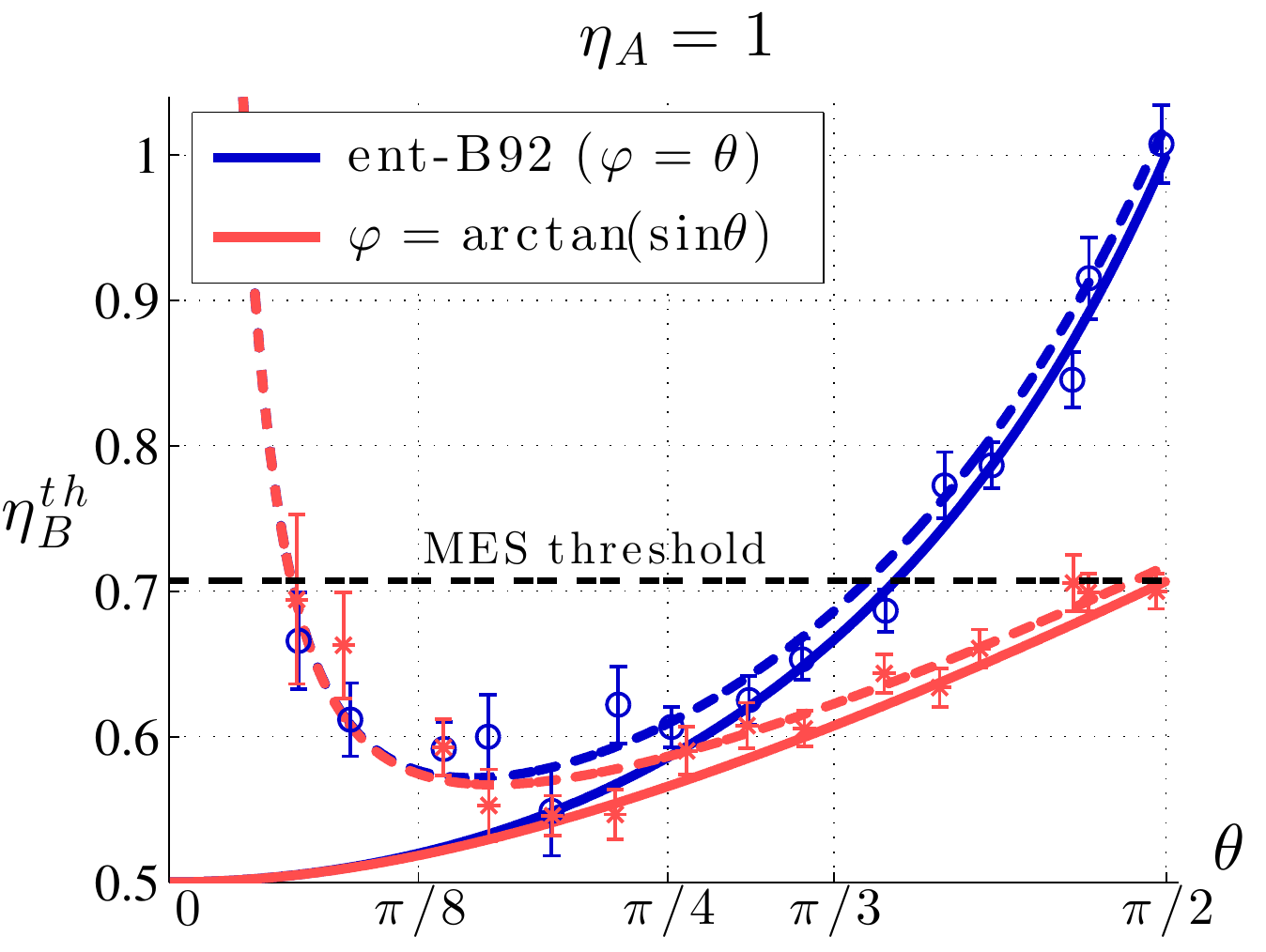}
\caption{Threshold detection efficiencies needed for
the violation of the Bell inequality in the $\eta_A=\eta_B=\eta$ case (left) and in the $\eta_A=1$ case (right).
Continuous and Dashed lines respectively 
represent theoretical values obtained in case of no imperfection and by using the noise model of eq. \eqref{model_appendix}.
We also report the threshold achievable with a perfect maximally entangled state (MES).}
\label{fig:eta}
\end{figure}

Let's consider the case $\eta_A=\eta_B=\eta^{th}$. From equation ({6}) of the main text, the threshold detection efficiency required to 
violate the Bell inequality can be predicted to be
\beq
\eta^{th}=\frac{p(a_1b_0)+p(a_1\bar b_0)+\frac12p(\bar a_0 b_1)+\frac12p(a_0b_1)+\frac12p(\bar a_0 b_0)+\frac12p(a_0b_0)}
{p(a_1b_1)+\frac12p(a_0b_1)+p(a_1b_0)-\frac12p(a_0b_0)+\frac12p(\bar a_0b_0)-\frac12p(\bar a_0b_1)}
\eeq
If $\eta_A=1$ (corresponding to Alice trusted device), the Bob's threshold detection efficiency $\eta_B$, can be be predicted to be
\beq
\eta^{th}_B=\frac{p(a_1b_0)+p(a_1\bar b_0)}{p(a_1b_1)+p(a_1b_0)-p(a_0b_0)-p(\bar a_0 b_1)}
\eeq
If Fig. \ref{fig:eta} we show the theoretical prediction and the  experimental values of $\eta^{th}$ and $\eta^{th}_B$.
Continuous lines represent theoretical prediction without imperfection, while dashed lines correspond to the prediction of the model
\eqref{model_appendix} with $p_w=0.007$ and $p_c=0.015$. Also in this case, we have a good agreement between the model
and the obtained results. As the model predicts, when white noise is turned on, states with very low entanglement cannot offer
advantages with respect to states with larger entanglement. In order to full exploit the properties of low entangled states 
it is necessary to generate quantum state with low amount of noise.

\end{document}